\documentclass{appolb}
\usepackage{graphicx}
\usepackage{cite}
\def\nn{\nonumber}

\newcommand{\secdec}{{\textsc{SecDec}}}
\newcommand{\secdecthree}{{\textsc{SecDec}~$3$}}
\newcommand{\pysecdec}{py{\textsc{SecDec}}}

\newcommand{\python}{{\texttt{python}}}
\newcommand{\form}{{\texttt{FORM}}}

\newcommand{\eps}{\epsilon}
\newcommand{\rd}{{\mathrm{d}}}

\begin{document}

\title{Numerical evaluation of two-loop integrals with pySecDec%
\thanks{Presented at the Final HiggsTools Meeting, IPPP, University of
Durham, UK, September 2017.}%
}
\author{
S.~Borowka, \address{Theoretical Physics Department, CERN, Geneva, Switzerland}\\[7mm]
{G.~Heinrich, S.~Jahn, S.P.~Jones, M.~Kerner,
\address{Max Planck Institute for Physics, F\"ohringer Ring 6, 80805 M\"unchen, Germany}}\\[7mm]
{J.~Schlenk 
\address{IPPP, University of Durham, South Road, Durham DH1 3LE, UK}}
}
\maketitle
\begin{abstract}
We describe the program \pysecdec{}, which factorises endpoint
singularities from multi-dimensional parameter integrals and can serve
to calculate integrals occuring in higher order perturbative
calculations numerically.  We focus on the new features and on
frequently asked questions about the usage of the program.
\end{abstract}

  
\section{Introduction}
At the CERN Large Hadron Collider (LHC), the exploration of the Higgs sector  has just begun. 
Data with unprecedented  precision are being and will be produced,
allowing us to further explore fundamental questions like 
the nature of electro-weak symmetry breaking, 
of which we only got a glimpse so far by the discovery of the Higgs
boson.  
The comparison of these data to theoretical predictions is vital in
order to identify effects of ``New Physics'', which may manifest
themselves first indirectly, via loop effects.
High precision  theoretical predictions are therefore mandatory for
the success of the LHC program, and even more so at
future colliders. 

To increase the precision of the theoretical predictions, 
higher orders in the perturbative expansion in the strong and
electroweak coupling constants need to be calculated. 
Such corrections often involve integrals depending on several
kinematic/mass  scales at two (or more) loops, where analytic results
are hard to achieve.

In these cases, numerical approaches may offer a solution. A method which proved useful 
in the presence of dimensionally regulated singularities is sector decomposition~\cite{Hepp:1966eg,Roth:1996pd,Binoth:2000ps,Heinrich:2008si}, 
as it provides an algorithm to factorise such singularities in an automated way.
The coefficients of the resulting Laurent series in the regulator
are  parametric integrals which can be integrated numerically.
This algorithm has been implemented in the program \secdec{}~\cite{Carter:2010hi,Borowka:2012yc,Borowka:2015mxa,Borowka:2017idc}, 
where from version 2.0~\cite{Borowka:2012yc} the restriction to Euclidean kinematics was lifted
by combining sector decomposition with a method to deform 
the multi-dimensional integration contour into the complex plane~\cite{Soper:1999xk,Beerli:2008zz}.
Other implementations of the sector decomposition algorithm can be found in
Refs.~\cite{Bogner:2007cr,Gluza:2010rn,Ueda:2009xx,Kaneko:2010kj,Smirnov:2008py,Smirnov:2009pb,Smirnov:2013eza,Smirnov:2015mct}. 

In this article, we describe the new version of the \secdec{}
program, called \pysecdec~\cite{Borowka:2017idc}. We particularly focus on the user
interface, providing answers to ``Frequently Asked
Questions''.


\section{Structure of the program}

The program consists of two basic parts: an algebraic part, 
based on \python{} and \form{}~\cite{Vermaseren:2000nd,Kuipers:2013pba,Ruijl:2017dtg},
and a numerical part, based on {\tt C++} code. 
The isolation of regulated endpoint singularities and the subsequent numerical integration 
can act on general polynomial functions, whereof Feynman integrals are a special case.

\begin{figure}[htb]
\begin{center}
\includegraphics[width=12cm]{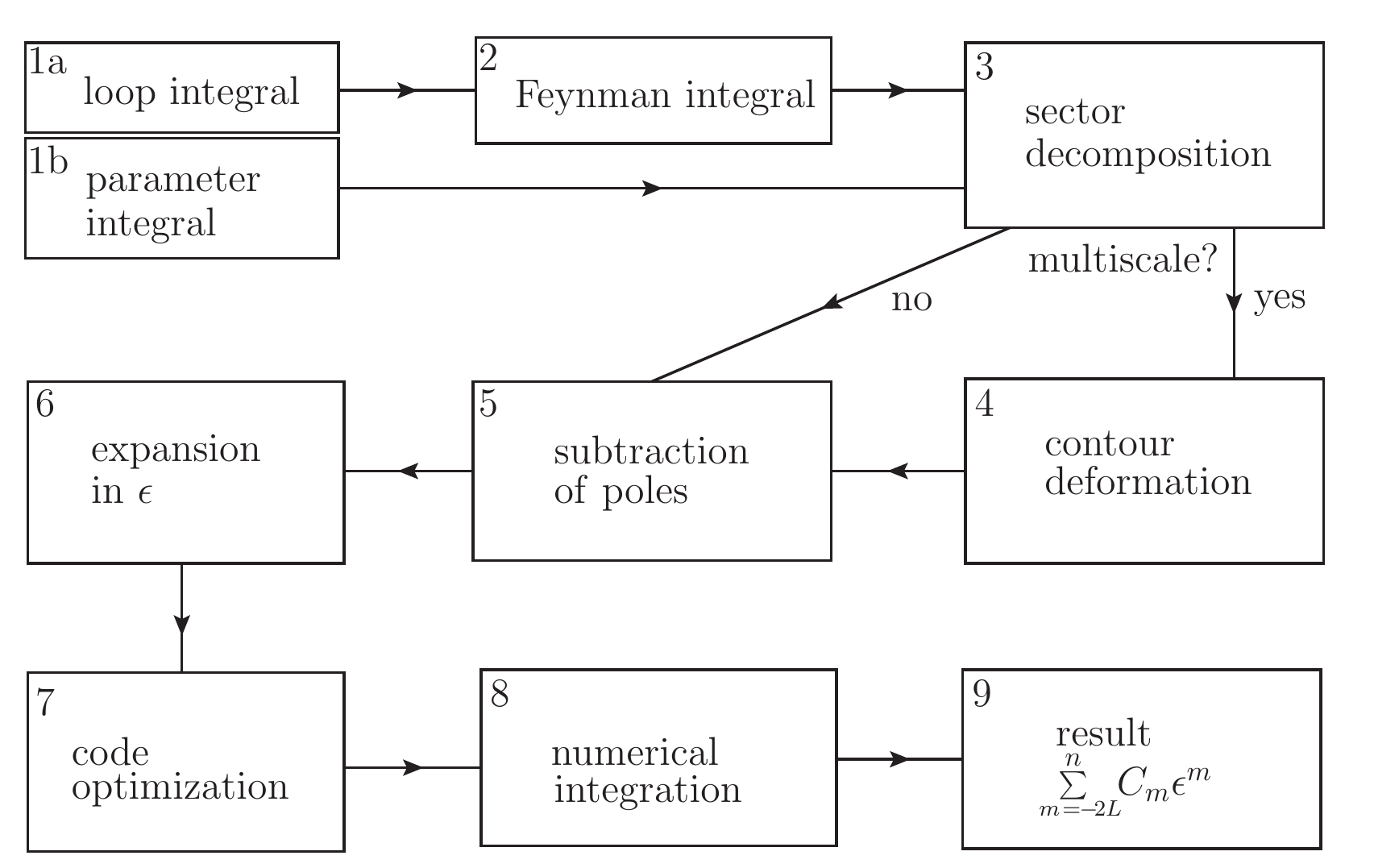}
\caption{Flowchart showing the main building blocks of
  \pysecdec{}. }
\label{fig:flowchart}
\end{center}
\end{figure}

Loop integrals (after Feynman parametrisation and momentum integration) 
can be considered as special cases of these more general polynomial integrands. 
In the \python{} code, this is reflected by the following structure: 
The \python{} function {\tt make\_package} accepts a list of polynomials raised to their individual powers as input -- corresponding to the box (1b) in Fig.~\ref{fig:flowchart}. 
In contrast, {\tt loop\_package} takes a loop integral (optionally
from either its graph or its propagator representation), which
corresponds to box (1a). After constructing the Feynman parametrisation of 
the loop integral, {\tt loop\_package} calls  {\tt make\_package} for further processing.
The steps (1) to (7) are performed in \python{} and \form{}, where \form{} produces 
optimized {\tt C++} code. The compiled integrand functions are by default combined into a library. For the numerical integration, 
we provide a simple interface to integrators from the {\sc Cuba}~\cite{Hahn:2004fe} library. The user also has direct access to the integrand functions, 
for example to pass them to an external integrator.

\subsection{Installation and usage}

The program can be downloaded from {\tt
  http://secdec.hepforge.org} or {\tt https://github.com/mppmu/secdec}.
It relies on  \python{} and runs with versions 2.7 and 3. It also uses
the packages {\tt numpy} ({\tt http://www.numpy.org}) and {\tt sympy}
({\tt http://www.sympy.org}). 

It is easiest to install \pysecdec{} from a release tarball available at
{\tt https://github.com/mppmu/secdec/releases/latest}. 
After download, \pysecdec{} is installed by the following shell commands:\\[3mm]
{\tt
tar -xf pySecDec-<version>.tar.gz \\
cd pySecDec-<version> \\
make \\
<copy the highlighted output lines into your .bashrc>
}

\medskip

The \texttt{make} command will automatically build further dependencies
in addition to \pysecdec{} itself. These are the
{\sc Cuba} library~\cite{Hahn:2004fe,Hahn:2014fua} needed for multi-dimensional numerical
integration, \form{}~\cite{Vermaseren:2000nd,Kuipers:2013pba,Ruijl:2017dtg} for the algebraic manipulation of
expressions and to produce optimized {\tt C++} code, {\sc
  Nauty}~\cite{2013arXiv1301.1493M} to find sector symmetries and the
GSL library~\cite{gsl}. 
The lines to be copied into the \texttt{.bashrc} define environment variables which make sure that
\pysecdec{} and its dependencies are found.
The \pysecdec{} 
user is strongly encouraged to cite the additional
dependencies when using the program.

\subsubsection*{Geometric sector decomposition strategies}
The program {\sc Normaliz}~\cite{2012arXiv1206.1916B,Normaliz} is needed for
the geometric decomposition strategies {\tt geometric} and {\tt geometric\_ku}.
In \pysecdec{}
version 1.3, the versions 3.0.0, 3.1.0, 3.1.1, 3.3.0 and 3.4.0 of {\sc Normaliz} are known to work.
Precompiled executables for different systems can be downloaded from \\
{\tt https://www.normaliz.uni-osnabrueck.de}. We recommend to
export its path to the environment of the terminal such that the
{\it normaliz} executable is always found. Alternatively, the path can be passed
directly to the functions that call it,  using {\tt normaliz\_executable=[path\_to\_normaliz]}.
The strategy {\tt iterative} can be used without having {\sc Normaliz} installed.

\subsection{Usage}
\label{subsec:usage}

The program comes with detailed documentation in both {\tt pdf}\\ ({\tt
  doc/pySecDec.pdf}) and {\tt html} ({\tt doc/html/index.html})
format. Online do\-cumentation can be found at  {\tt https://secdec.readthedocs.io/en/latest}.
In the {\tt examples} folder we provide examples  for several ways how to apply the
program. One is to use \pysecdec{} in a ``standalone" mode to obtain numerical results for individual integrals. This
corresponds to a large extent to the way previous \secdec{} versions were used.
The other allows the generation of a library which can be linked to the calculation of amplitudes or
other expressions, to evaluate the integrals contained in these expressions.

To get started, we recommend to read the section ``getting started" in the online documentation.
The basic steps can be summarised as follows:
\begin{enumerate}
\item Produce a \python{} script to define the integral, the replacement rules for the kinematic invariants,
the requested order in the regulator and some other options\\ 
(see e.g. the one-loop box example {\tt box1L/generate\_box1L.py}).
\item Run the script using \python{}. This will generate a
  subdirectory according to the {\tt name} specified in the script.
\item Type {\tt make -C <name>}, where {\tt <name>} is your chosen name. This will create the {\tt C++} libraries.
\item Produce a \python{} script to perform the numerical integration using the
  \python{} interface (see e.g. {\tt box1L/integrate\_box1L.py}).
\end{enumerate}
Further usage options such as looping over multiple kinematic points
are described in the documentation and in Ref.~\cite{Borowka:2017idc}.

\vspace*{5mm}

The algebra package can also be used for symbolic manipulations on integrals. This can be
of particular interest when dealing with non-standard loop integrals, or if the user would like
to interfere at intermediate stages of the algebraic part.

\subsection{New features}

In addition to the complete re-structuring and usage of open source
software only, there are various new features compared to
\secdecthree{}: 
\begin{itemize}
\item The functions can have any number of regulators for endpoint singularities, not
  only the dimensional regulator $\eps$.
\item The treatment of numerators of loop integrals is more flexible. Numerators can
  be defined in terms of contracted Lorentz vectors or inverse
  propagators or a combination of both.
\item The distinction between ``general functions" and ``loop
  integrands" is removed in the sense that all features are available
  for both, loop integrals and general polynomial functions (as far as
  they make sense outside the loop context). 
\item The inclusion of additional functions which do not enter the decomposition
  has been facilitated and extended.
\item The treatment of poles which are higher than logarithmic has
  been improved.
\item A procedure has been implemented to detect and remap
  singularities at $x_i=1$ which result from special kinematic configurations.
\item A symmetry finder~\cite{Jones2017:ACAT} has been implemented which can detect isomorphisms between sectors.
\item Diagrams can be drawn (optionally, based on {\tt neato}~\cite{graphviz}; the program will however run normally if {\tt neato} is not installed).
\item The evaluation of multiple integrals or even amplitudes is now possible, using the generated {\tt C++} library.
\end{itemize}

\subsection{Frequently asked questions}

In the following we list some questions which may come up during usage
of the program, and  give  answers which should spare the user to search the manual.

\begin{itemize}
\item How can I adjust the numerical integration parameters?

If the python interface is used for the numerical integration,
i.e.  a python script like  {\tt examples/integrate\_box1L.py},
the integration parameters can be specified in the argument list of
the integrator call. For example, using {\tt Vegas} as integrator: \\
{\tt
  box1L.use\_Vegas(flags=2, epsrel=1e-3, epsabs=1e-12,}\\
{\tt nstart=5000, nincrease=10000, maxeval=10000000, }\\
{\tt real\_complex\_together=True) }

Or, using {\tt Divonne} as integrator:\\
{\tt box1L.use\_Divonne(flags=2, epsrel=1e-3,
  epsabs=1e-12, }\\
{\tt maxeval=10000000, border=1e-8, real\_complex\_together=True) }

The parameter {\tt real\_complex\_together} tells the integrator to
integrate real and imaginary parts simultaneously. 
A list of possible options for the integrators can be found at the end
of Section 5.9 of the manual.

\item How can I increase the numerical accuracy?

The integrator stops if any of the folllowing conditions is
fulfilled: (1) {\tt epsrel} is reached, (2) {\tt epsabs} is reached,
(3) {\tt maxeval} is reached.
Therefore, setting these parameters accordingly will cause the
integrator to make more iterations to reach a more accurate result.

\item How can I tune the contour deformation parameters?

You can specify the parameters in the argument of the integral call in
the python script for the integration, see e.g. line 12 of \\
 {\tt examples/integrate\_box1L.py}:
 
{\small 
{\tt str\_integral\_without\_prefactor, str\_prefactor,}\\
{\tt str\_integral\_with\_prefactor=box1L(real\_parameters=[4.,-0.75,1.25,1.],}\\
{\tt number\_of\_presamples=1000000,deformation\_parameters\_maximum=0.5)}
}

This sets the number of presampling points to $10^6$ (default: $10^5$)
and the maximum value for the contour deformation parameter $\lambda$,
{\small {\tt deformation\_parameters\_maximum}}, to 0.5 (default: 1). The user should make sure that
{\small {\tt deformation\_parameters\_maximum}} is always larger than {\small {\tt deformation\_parameters\_minimum}} (default: $10^{-5}$).
These parameters are explained in Section 5.9. of the manual  under
``Parameters''.

\item What can I do if the program stops with an error message
  containing {\it ``sign\_check\_error''}?

This error occurs if the contour deformation leads to a wrong sign of
the Feynman $i\,\delta$ prescription, usually due to the fact that the
deformation parameter $\lambda$ is too large.

Choose a larger value for {\tt number\_of\_presamples}  and a smaller
value (e.g. 0.5) for {\tt deformation\_parameters\_maximum} (see item above).
If that does not help, you can try 0.1 instead of 0.5 for \\
{\tt  deformation\_parameters\_maximum}.

\item What does ``additional\_prefactor'' mean exactly?

We should first point out that the conventions for additional prefactors defined by the user have been changed
between \secdecthree{} and \pysecdec{}. The prefactor specified by the
user will now be {\it included} in the numerical  result.

To make clear what is meant by ``additional'', we repeat our
conventions for Feynman integrals here:\\
A scalar Feynman graph $G$ in $D$ dimensions 
at $L$ loops with  $N$ propagators, where 
the propagators can have arbitrary, not necessarily integer powers $\nu_j$,  
has the following representation in momentum space:
\begin{eqnarray}\label{eq0}
G &=& \int\prod\limits_{l=1}^{L} \rd^D\kappa_l\;
\frac{1}
{\prod\limits_{j=1}^{N} P_{j}^{\nu_j}(\{k\},\{p\},m_j^2)}\nn\\
\rd^D\kappa_l&=&\frac{\mu^{4-D}}{i\pi^{\frac{D}{2}}}\,\rd^D k_l\;,\;
P_j(\{k\},\{p\},m_j^2)=(q_j^2-m_j^2+i\delta)\;,
\end{eqnarray}
where the $q_j$ are linear combinations of external momenta $p_i$ and loop momenta $k_l$.
Introducing Feynman parameters leads to 
\begin{eqnarray}\label{EQ:param_rep}
G &=& (-1)^{N_{\nu}}
\frac{\Gamma(N_{\nu}-LD/2)}{\prod_{j=1}^{N}\Gamma(\nu_j)}\int
\limits_{0}^{\infty} 
\,\prod\limits_{j=1}^{N}dx_j\,\,x_j^{\nu_j-1}\,\delta(1-\sum_{l=1}^N x_l)\,\frac{{\cal U}^{N_{\nu}-(L+1) D/2}}
{{\cal F}^{N_\nu-L D/2}} \nonumber\\
\end{eqnarray}
The prefactor $ (-1)^{N_{\nu}}\,\Gamma(N_{\nu}-LD/2)/\prod_{j=1}^{N}\Gamma(\nu_j)$ coming from the Feynman parametrisation
will always be included in the numerical result, corresponding to {\tt
  additional\_prefactor=1} (default), i.e. the program will return the
numerical value for $G$.
If the user defines {\tt additional\_prefactor=`gamma(3-2*eps)'}, this prefactor will be expanded in $\eps$
and included in the numerical result returned by \pysecdec{}, in
addition to the one coming from the Feynman parametrisation.

For general polynomials not related to loop
integrals, i.e. in {\small {\tt make\_package}}, the  prefactor provided by the
user is the only prefactor, as there is no prefactor coming from a
Feynman parametrisation in this case. This is the reason why in {\tt make\_package}
the keyword for the prefactor defined by the user 
is  {\tt prefactor}, while in  {\tt loop\_package} it is {\tt additional\_prefactor}. 

\item What can I do if I get `nan'?

This means that the integral does not converge and can have several
reasons. 
When Divonne is used as an integrator, it is important to use a
non-zero value for {\tt border}, e.g. {\tt border=1e-8}.
Vegas is in general the most robust integrator.
When using Vegas, try to increase the values for {\tt nstart} and {\tt
  nincrease}, for example {\tt nstart=10000} (default: 1000) and {\tt
  nincrease=5000} (default: 500).

\item Can I include my own functions in the numerator of a loop
  integral?

Yes, as long as the functions are finite in the limit $\eps\to 0$. 
The nu\-me\-rator should be a sum of products of  numbers,
scalar products (e.g. {\tt `p1(mu)*k1(mu)*p1(nu)*k2(nu)'} and/or symbols
(e.g. {\tt `m'}).
 The default numerator is 1.
Examples:

{\tt p1(mu)*k1(mu)*p1(nu)*k2(nu) + 4*s*eps*k1(mu)*k1(mu)}\\
{\tt p1(mu)*(k1(mu) + k2(mu))*p1(nu)*k2(nu)}\\
{\tt p1(mu)*k1(mu)*my\_function(eps)}

More details can be found in section 5.2.1 of the manual.

\item How can I integrate just one coefficient of a particular order
  in $\eps$\,?

You can pick a certain order  in the
{\tt C++} interface (see  section 2.2.4 of the manual).
To integrate only one order, change the line

{\small 
{\tt  const box1L::nested\_series\_t<secdecutil::UncorrelatedDeviation}\\
{\tt <box1L::integrand\_return\_t>> result\_all = secdecutil::}\\
{\tt deep\_apply( all\_sectors, integrator.integrate );}
}

to

{\small 
{\tt     int order = 0; } {\it // compute finite part only}\\
{\tt     const secdecutil::UncorrelatedDeviation<box1L::integrand\_return\_t> }\\
{\tt result\_all = secdecutil::deep\_apply( }\\
{\tt all\_sectors.at(order), integrator.integrate );}
}

where {\tt box1L} is to be replaced by the name of your integral.
In addition, you should remove the lines
\begin{center}
\includegraphics[width=11cm]{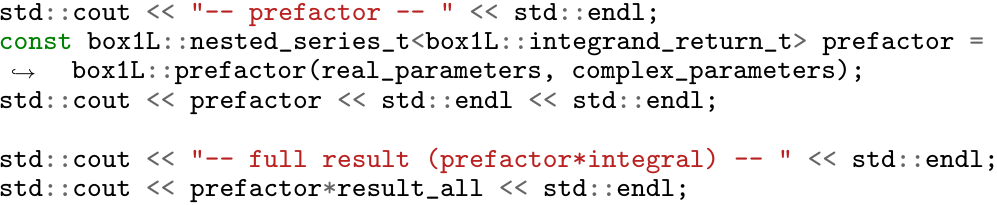}
\end{center}


because the expansion of the prefactor will in general mix with the pole
coefficients and thus affect the finite part. 
We should point out however that deleting these lines also means that the result will not contain any prefactor, 
not even the one coming from the Feynman parametrisation.

\item How can I use complex masses?

In the python script generating the expressions for the integral,
define {\tt  mass\_symbols} in the same way as for real masses, e.g.

{\tt Mandelstam\_symbols = ['s', 't']}\\
{\tt mass\_symbols = ['msq']}

In {\tt loop\_package} then define 

{\tt real\_parameters = Mandelstam\_symbols,}\\
{\tt complex\_parameters = mass\_symbols,}\\
{\tt . . .}\\

In the integration script (using the python interface), 
the numerical values for the complex parameters are given after the
ones for the real parameters:

{\tt str\_integral\_without\_prefactor, str\_prefactor, }\\ 
{\tt str\_integral\_with\_prefactor = integral(}\\
{\tt real\_parameters=[4.,-1.25],complex\_parameters=[1.-0.0038j])}\\

Note that in python the letter `j' is used rather than `i' for the imaginary part.

\item When should I use the ``split'' option?

This option can be useful for integrals which do not have a Euclidean region. 
If certain kinematic conditions are fulfilled, for example if the integral contains massive on-shell lines,
it can happen that singularities at $x_i=1$ remain in the ${\cal F}$ polynomial after the decomposition.
The split option remaps these singularities to the origin of parameter space. 
If your integral is of this type, and with the standard approach the numerical integration does not seem to converge, 
try the ``split" option. It produces a lot more sectors, so it should not be used without need.

We also would like to mention that very often a change of basis can be beneficial if integrals of this type occur in the calculation.

\end{itemize}

\section{Conclusions}
\label{sec:conclusion}
We have described new features of the program \pysecdec{},  which is publicly available at {\tt https://github.com/mppmu/secdec}.
\pysecdec{} is entirely based on open source software. 
The algebraic part can isolate end-point singularities in any number of regulators from general polynomial expressions,
for example multi-loop Feynman integrals.
For the numerical part, a library of {\tt C++} functions is created, which allows very flexible usage, 
and in general outperforms \secdecthree{} in the numerical evaluation times.
In particular, it extends the functionality of the program from the evaluation of individual (multi-)loop integrals  
to the evaluation of larger expressions containing multiple integrals, as for example two-loop amplitudes. 
We have also provided answers to some questions which were frequently
asked by \pysecdec{} users.

\section*{Acknowledgements}
This research was supported in part by the 
Research Executive Agency of the European Union under the Grant Agreement
PITN-GA2012316704 (HiggsTools) and by the Munich Institute for Astro- and Particle Physics (MIAPP) of the
DFG cluster of excellence ``Origin and Structure of the Universe".

 
\bibliographystyle{JHEP}

\bibliography{refs_secdec}

\end{document}